# Thermals Forecast with Temps

Oliver Predelli
Braunschweig, Germany
segelflieger@e-mail.de

**Can you read the strength of thermals and their progression in altitude from a Temp? Yes, it works! So the Skew-T log-P diagram or emagram (called "Temp") is a central element of weather forecasting for gliding and soaring.**

Sometimes I envy club members who can take to the air whenever the "weather" is right. I, however, have family and work commitments that allow me to fly only a few days a year. In our kitchen calendar, these days are planned well in advance. Rescheduling a flying day is not easy.

That's why I need a reliable weather forecast. If possible, I would like to know several days in advance what weather to expect. Will it be a cross-country flight, or will it only be enough for aerodrome rounds? I also don't want to be disappointed by wrong forecasts. How often has it happened that the gliding weather forecast promised good thermals the day before, but on the morning of the flight day it looked rather dull? So it would be quite helpful to supplement the official flying weather forecast with your own forecast.

Besides weather charts for barometric pressure, geopotential altitude, or equivalent potential temperature, a Temp is a common tool for making your own weather predictions. These vertical profiles, these horizontal slices through the atmosphere, are an elegant way to figure out what the weather will be like when gliding. It is well known that temperatures, base altitude, cloud cover, and risk of overdevelopment can be read from an emagram [Haf99]. But does the diagram also give information about the strength of the thermal, i.e. the updraft velocity? To answer this question, one must delve a little deeper into the nature of thermals.

Of course, every professional gliding weather report gives a value for the expected thermal strength. At DWD, TopMeteo and RASP this value is a central part of the forecasts. The weather services also need it for estimating the possible flight distance over the day. A formula used for this, at least at RASP, is $w^* = ((g/T_0) \cdot Q_S \cdot D)^{1/3}$ [Blu17].

In deriving this formula, it was assumed that there is always a temperature difference between thermals and ambient air [All06]. Thermals, it is said, result from the fact that the air on the ground heats up, separates at a certain point and retains a certain temperature advantage over the ambient air during the ascent. For a rough calculation, this reasoning may give good results, but it does not quite correspond to reality. This is because the thermal strength does not depend on how much solar energy has previously reached the ground and how high the convection reaches. Rather, it depends solely on the buoyancy forces that drive it upward after detachment. The density differences between thermals and ambient air play a crucial role. As we now know, the temperature differences between thermals and ambient air are negligible [Blu14]. Therefore, the temperature difference can hardly be relevant for the density difference. Instead, the density depends significantly on the relative humidity, i.e., the dew point depression. The dew point temperature of thermal air is higher than the dew point temperature of ambient air. The air temperature and both dew point temperatures can be read from a Temp. So you should also be able to determine the updraft velocity from a Temp. And this for the respective flight altitude.

### The thermal formular

After some theoretical considerations and several mathematical transformations, a simple equation for the updraft velocity can indeed be found that can be applied to the Temp (see the appendix at the end of this article):

$$w \approx K \cdot \sqrt{\frac{1{,}1^{(\tau_{Th}-\tau_{Lu})/°C}-1}{1{,}1^{(\vartheta_{Lu}-\tau_{Lu})/°C}}} \quad \text{with } K = 5{,}6 \text{ m/s.}$$

Where $\vartheta_{Lu}$ is the temperature of the ambient air, $\tau_{Lu}$ is the dew point of the ambient air, and $\tau_{Th}$ is the dew point of the rising thermal, each at the corresponding height above ground. The difference $(\vartheta_{Lu} - \tau_{Lu})$ is nothing but the spread, and $(\tau_{Th} - \tau_{Lu})$ is the temperature difference between the dew point of the ambient air and the dew point of the thermal.

The thermal formula proves: The strength of thermals is largely determined by differences in humidity. The drier the ambient air and the more humid the rising air, the stronger the thermal.

### Using the thermal formula

The Temp shows the respective ambient air temperature ($\vartheta_{Lu}$) and dew point temperature ($\tau_{Lu}$) at different heights of the convective layer. Furthermore, we know that the dew point temperature in the rising air follows the line of the constant saturation mixing ratio of the ground dew point, also shown in the emagram. Thus, on this line lies the dew point temperature of the thermal ($\tau_{Th}$) at the respective height.

We only need to read the three values $\vartheta_{Lu}$, $\tau_{Lu}$ and $\tau_{Th}$ from the Temp relevant for our flight region and can use the above formula to calculate the strength of the thermal at different altitudes. **Fig. 1** shows an example of this using a Temp from the GFS visualization software zyGrib (www.zygrib.org). On July 7, 2016, an average climb of 2.1 m/s was to be expected

---

at an altitude of 1,000 m above Wilsche landing field, but the aircraft's own descent must still be subtracted from this.

Instead of pulling out the calculator every time, you can create a small table for the relevant temperature values (**Fig. 2**). With a little practice, you will have the typical gradient values in your head after a short time. All I do to estimate the thermal conditions is look at how the spread changes and where the spread is approximately divided by the line of the saturation mixing ratio of the ground pressure.

**Parameter identifikation**
I would also like to explain how I determined the parameter K = 5.6 m/s. Instead of calculating it elaborately from the various physical quantities in the equations I used, it was easier to determine it by measurement. After all, many of these physical quantities are difficult to quantify. In any case, I don't know the $c_w$ value of a rising thermal bubble. But how should a measurement work? Quite simply: by analyzing real flight data! There are databases on the Internet from which you can read how strong the thermals were at different heights of the convective layer at different locations. We glider pilots all know them, they are http://www.onlinecontest.org or https://skylines.aero/ with their thousands of logger files.

No, of course it is not the true thermal strength that is displayed there in m/s when circling in the updraft. The inherent sinking of the glider must be taken into account. The pilot may not have centered the thermal properly or his circling may have been poorly executed. Also, locally there may have been a decidedly stronger or weaker updraft. However, by a statistical evaluation of many flights the possible errors balance out and an average value can be recognized.

First, I searched the University of Wyoming website http://weather.uwyo.edu/upperair/sounding.html for suitable sites for radiosonde ascent. Finally, verification of the thermal formula should be based on real weather data, not on GFS model data. The weather balloon ascents should occur in a region where glider pilots are out and about posting their logger files on the Internet. They should also take place at a time when gliders are in the air, so that one can put the logger records in a temporal context with the balloon measurements. For my evaluations I used in particular the data from Bergen near Celle, Lindenberg and Kümmersbruck (all in Germany), as well as from Trappes (France), Legionowo (Poland) and Tuscon (USA). The data from Australia, on the other hand, were not useful because the balloons there ascend around 9 a.m. and 9 p.m. local time, i.e. when the thermals have just started or have already passed.

From the radiosonde data, the theoretical climb values were calculated using the thermal formula and compared with the climb values of the flights passing there. Using mathematical parameter identification, the missing value for K could be determined. A comparison between the calculated values and the climb values of the gliding flights shows a good correlation (**Fig. 3**).

**Fig. 4** and **Fig. 5** show this evaluation exemplarily for a flight that took me around Bergen (Germany) at the right time on July 7, 2016. In contrast to Fig. 1, which shows GFS model data, Fig. 5 uses radiosonde values. This explains the slight differences between the two diagrams. The green curve shows the strength of the thermal, which is calculated from the air temperature (blue), the dew point of the ambient air at the respective altitude (red), and the line of the constant saturation mixing ratio (orange). The cloud base is at about 1,200 m. The dew point difference of 2.5 °C at the inversion suggests an occultation of 7 eighths. The calculated thermal shows 2 m/s. And so it was in the area that day. The sky became more and more cloudy. The thermals were rather moderate.

Boom weather is different. June 11, 2016 was such a day, with more than 10,000 route kilometers covered from Wilsche alone, including two flights of more than 1,000 km. In Fig. 6 we see the GFS Temp for the area of Wittenberge on the Elbe (Germany). Theoretically, at 15 UTC, 2.9 m/s were still expected at 1,000 m altitude.

**Limits of the procedure**
The forecast stands and falls with the quality of the temperature forecast. For example, if the GFS model is inaccurate, the result of the thermal formula will also be incorrect. Of course, the temperature pattern must allow for thermals. If there is a ground inversion, thermals cannot develop and the method is not applicable.

Locally, thermals can be stronger or weaker depending on the soil type (temperature) and moisture of the soil. In addition, the method cannot show how far apart the thermal updrafts are.

In the lee of lakes or forest edges, the thermal is sometimes more humid, and the climb values are better than the calculated average. This moister air on the ground shifts the line of constant saturation mixing ratio in the emagram to the right. With the above considerations, it is immediately clear that this must increase the thermal strength.

In the mountains, thermals intensify on the sunny side of the slopes; here the calculated climb rates must be adjusted.

The inherent sink rate must be subtracted from the calculated thermal strength if you want to estimate your climb rate in the glider. However, whether a pilot can actually use the thermal potential depends primarily on his flying skills.

The discrepancies between theory and practice shown in Fig. 3 need not surprise us. I even find it reassuring that, despite all the mathematics, a glider pilot still has to seek his luck himself with stick in hand.

**Where to get suitable Temps?**
My aviation weather forecast for the next days is based on the data of the "Global Forecast System" (GFS) of the US National Oceanic and Atmospheric Administration (NOAA) available on the Internet. For visualization I use "XyGrib" (https://opengribs.org). The open-source software XyGrib can generate forecast maps from the GFS data not only for a specific region, but also forecast temperatures for arbitrary locations.

RASP calculates good Temps in principle, but their scaling is a bit unwieldy for our method. Another source is http://www.weatheronline.co.uk/cgi-bin/expertcharts, which provides a link to the GFS forecast Temps at the top right of the web page.

However, a Temp with data from a radiosonde ascending at night is completely inappropriate. We are familiar with the procedure of plotting the expected temperature gradients of the day on such graphs to estimate initial and base thermal heights [Haf99]. However, since we cannot use this to find out how the dew point temperatures at different heights change during the day, we lack a key input variable for our thermal formula.

**Summary**
In this article, the thermal formula was presented for the first time. With it, any glider pilot can estimate how strong the thermals will be based on forecast temperatures, e.g. based on GFS model data. And, this is also new, one can immediately see in which altitude band the thermals are best. If software such as XyGrib is used, these forecasts can be made for any location on earth several days in advance - free of charge and with great accuracy.

**Appendix**
The following assumption: The thermal has the form of a "bubble". There is no gas exchange with the environment [Haf99, p. 12ff]. It is lighter than the ambient air and rises from the ground like a gas balloon. Volume and shape of the thermal bubble are sufficiently constant during the ascent. The change in the rate of ascent is negligibly small.

Buoyancy, weight and drag forces act on the thermal bubble in equilibrium:

$$F_A = F_G + F_W \quad , \qquad (1)$$

with

$$F_A = \rho_{Lu} \cdot V_{Th} \cdot g \quad , \qquad (2)$$
$$F_G = \rho_{Th} \cdot V_{Th} \cdot g \quad , \qquad (3)$$
$$F_W = \frac{1}{2} \cdot c_W \cdot A_{Th} \cdot \rho_{Lu} \cdot w^2 \quad , \qquad (4)$$

By substitution, transformation and introduction of the constant parameter

$$A = \frac{2 \cdot g \cdot V_{Th}}{c_W \cdot A_{Th}} \qquad (5)$$

the equation for the buoyancy velocity is obtained:

$$w = \sqrt{A \cdot \frac{\rho_{Lu} - \rho_{Th}}{\rho_{Lu}}} \quad . \qquad (6)$$

The buoyancy velocity w is thus calculated directly from the density of the ambient air and the density of the air inside the thermal bubble.

According to [Hak16, p. 40], the air density ρ is calculated from

$$\rho = \frac{p}{R_f \cdot T} \quad , \qquad (7)$$

there is

$$R_f = \frac{R_t}{1 - \varphi \cdot p_d/p \cdot (1 - R_t/R_d)} \quad , \qquad (8)$$

with $R_t$ = 287,05 J/kg·K as the gas constant for dry air and Rd = $R_d$ = 461 J/kg·K as the gas constant for water vapor.

Eq. 8 is combined with

$$a = p_d/p \cdot (1 - R_t/R_d) \qquad (9)$$

to give

$$R_f = \frac{R_t}{1 - a \cdot \varphi} \quad . \qquad (10)$$

Here a simplification of the further calculation takes place, because the term $p_d/p$ is actually temperature and altitude dependent. The factor $a$ determined by the saturation vapor pressure and the atmospheric pressure is the same for the ambient air and the thermal, because both have approximately the same temperature at the respective altitude [Blu14, S. 10ff].

Substituting Eq. 7 and Eq. 10 into Eq. 6, we obtain after some transformations

$$w^2 = A \cdot \frac{\varphi_{Th} - \varphi_{Lu}}{B - \varphi_{Lu}} \quad . \qquad (11)$$

with $\quad A = \dfrac{2 \cdot g \cdot V_{Th}}{c_W \cdot A_{Th}} \quad$ and $\quad B = \dfrac{1}{a} \quad$ .

The dew point temperature τ is calculated from the relative humidity φ and the air temperature ϑ according to [Int01]

with $K_2$ = 22,46 and $K_3$ = 272,62 °C:

$$\tau = K_3 \cdot \frac{\frac{K_2 \cdot \vartheta}{K_3 + \vartheta} + \ln \varphi}{\frac{K_2 \cdot K_3}{K_3 + \vartheta} - \ln \varphi} \quad . \quad (12)$$

Eq. 12 can be rearranged to φ:

$$\ln \varphi = \frac{K_2 \cdot K_3 \cdot (\tau - \vartheta)}{(K_3 + \vartheta) \cdot (K_3 + \tau)} \quad . \quad (13)$$

Substituting Eq. 13 into Eq. 11 to calculate both the relative humidity of the ambient air and the relative humidity of the thermal bubble yields the **equation for calculating the thermal strength**:

$$w = \sqrt{A \cdot \frac{e^{\frac{K_2 \cdot K_3 \cdot (\tau_{Th} - \vartheta_{Lu})}{(K_3 + \vartheta_{Lu}) \cdot (K_3 + \tau_{Th})}} - e^{\frac{K_2 \cdot K_3 \cdot (\tau_{Lu} - \vartheta_{Lu})}{(K_3 + \vartheta_{Lu}) \cdot (K_3 + \tau_{Lu})}}}{B - e^{\frac{K_2 \cdot K_3 \cdot (\tau_{Lu} - \vartheta_{Lu})}{(K_3 + \vartheta_{Lu}) \cdot (K_3 + \tau_{Lu})}}}} \quad (14)$$

Because of the relatively high value of $K_3$ = 272,62 °C, Eq. 13 can be simplified, accepting some error, with ($K_3 + \vartheta$) ≈ ($K_3 + \tau$) ≈ $K_3$ to be

$$\ln \varphi \approx \frac{K_2}{K_3} \cdot (\tau - \vartheta) \quad . \quad (15)$$

From this follows with the values for $K_2$ and $K_3$ from Eq. 12:

$$\varphi \approx e^{\frac{K_2}{K_3} \cdot (\tau - \vartheta)} \approx 1,1^{(\tau - \vartheta)/°C} \quad . \quad (16)$$

Converting the (τ -ϑ) into -(ϑ -τ) and substituting Eq. 16 into Eq. 11, we obtain:

$$w^2 = A \frac{\cdot 1,1^{(\tau_{Th} - \vartheta_{Lu})/°C} - 1,1^{(\tau_{Lu} - \vartheta_{Lu})/°C}}{B - 1,1^{(\tau_{Lu} - \vartheta_{Lu})/°C}} \quad . \quad (16)$$

Dividing the numerator and denominator by $1,1^{(\tau_{Lu} - \vartheta_{Lu})/°C}$ and eliminating B because 1/B << 1 gives the **simplified thermal formula**:

$$w \approx K \cdot \sqrt{\frac{1,1^{(\tau_{Th} - \tau_{Lu})/°C} - 1}{1,1^{(\vartheta_{Lu} - \tau_{Lu})/°C}}} \quad . \quad (17)$$

with $\quad K = \sqrt{\frac{A}{B}} = \sqrt{\frac{2 \cdot a \cdot g \cdot V_{Th}}{c_W \cdot A_{Th}}} \quad .$

**Formula symbols**

- $F_A$   Buoyancy force of the thermal bubble in N
- $F_G$   Weight force of the thermal bubble in N
- $F_W$   Resistance force counteracting the upward movement of the thermal bubble in N
- $ρ_{Lu}$   Density of the ambient air in kg/m³
- $ρ_{Th}$   Density of the air inside the thermal bubble in kg/m³
- $g$   Acceleration due to gravity in m/s²
- $V_{Th}$   Volume of the thermal bubble in m³
- $A_{Th}$   Cross-sectional area of the thermal bubble relevant for the drag force in m²
- $c_w$   Drag coefficient of the rising thermal bubble
- $w$   Updraft velocity of the thermal bubble in m/s
- $p$   Ambient pressure in Pascal
- $p_d$   Saturation vapor pressure in Pascal
- $T$   Temperature in Kelvin
- $\vartheta_{Lu}$   Temperature of ambient air in °C
- $R_f$   Gas constant for humid air in J/kg·K
- $\varphi_{Lu}$   relative humidity of ambient air (e.g. 0.8 for 80%)
- $\varphi_{Th}$   relative humidity in the thermal bubble (e.g. 0.8 for 80%)
- $\tau_{Lu}$   Dew point temperature of the ambient air in °C
- $\tau_{Th}$   Dew point temperature of the thermal bubble in °C
- $w^*$   Updraft velocity for a given terrain section in m/s
- $T_0$   Daily mean temperature in K
- $Q_S$   Energy input of the sun in W/m²
- $D$   Height of the thermally mixed layer in m

**Literature**

[All06] Allen, M. J.: "Updraft Model for Development of Autonomous Soaring Uninhabited Air Vehicles", 44th AIAA Aerospace Sciences Meeting and Exhibit, Reno, 2006

[Blu14] Blum, H.: "Meteorologie für Segelflieger", Motorbuch Verlag, Stuttgart, 2014

[Blu17] Blum, H.: "Modell V(orh)ersagen", segelfliegen 1/2017, Gabler Media, Bilten GL, 2017

[Haf99] Hafner, T.: "Handbuch der Flugwettervorhersagen für den Luftsport", Preprint to the 26. World Gliding Championships 1999 in Bayreuth

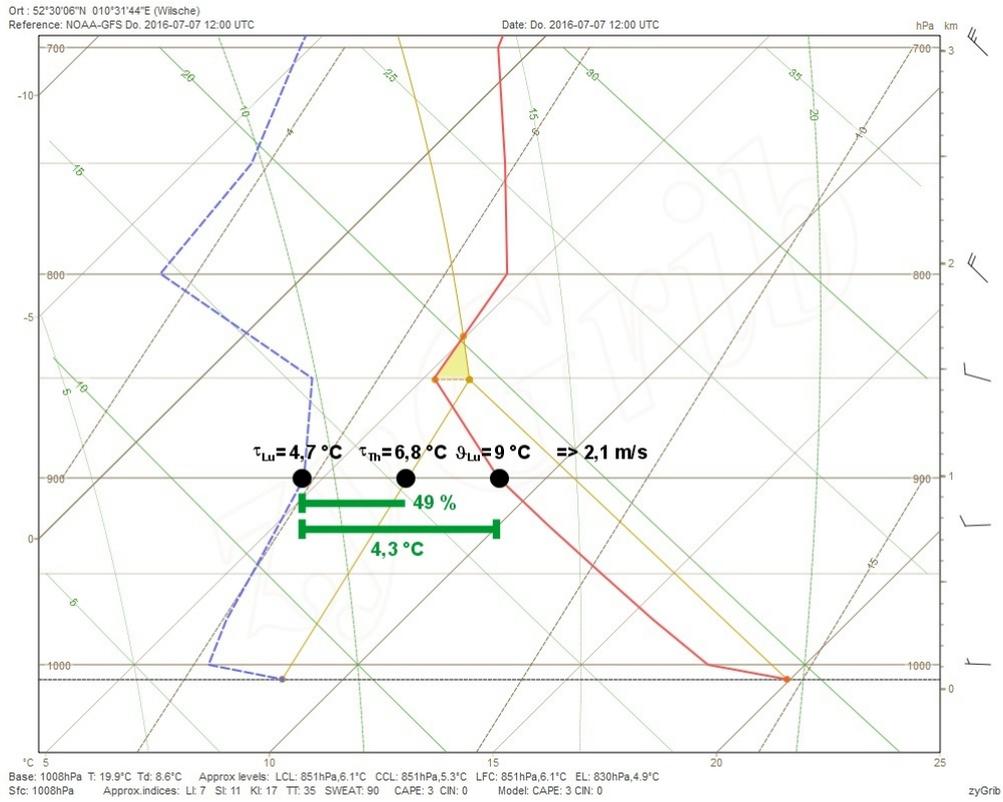

**Fig. 1**: Temperature values for calculating the thermal strength for July 7, 2016 in Wilsche (Germany). Source of the Temp: zyGrib.

| Thermal updraft velocity in m/s | | | | |
|---|---|---|---|---|
| Mixing ratio of lifted parcel divides spread at | | | | |
| Dew-point spread | 25 % | 50 % | 75 % | 100 % |
| 2 °C | 1,1 | 1,6 | 2,0 | 2,3 |
| 4 °C | 1,5 | 2,1 | 2,7 | 3,2 |
| 6 °C | 1,6 | 2,4 | 3,1 | 3,7 |
| 8 °C | 1,8 | 2,6 | 3,4 | 4,1 |
| 10 °C | 1,8 | 2,7 | 3,6 | 4,4 |
| 12 °C | 1,8 | 2,8 | 3,7 | 4,6 |
| 14 °C | 1,8 | 2,8 | 3,8 | 4,8 |
| 16 °C | 1,8 | 2,8 | 3,8 | 5,0 |
| 18 °C | 1,7 | 2,8 | 3,8 | 5,1 |
| 20 °C | 1,7 | 2,7 | 3,8 | 5,2 |
| 22 °C | 1,6 | 2,7 | 3,8 | 5,2 |
| 24 °C | 1,6 | 2,6 | 3,8 | 5,3 |
| Average | 1,5 | 2,5 | 3,5 | 4,5 |

**Fig. 2**: Simplified table for direct reading of thermal updraft velocities based on temperature differences.

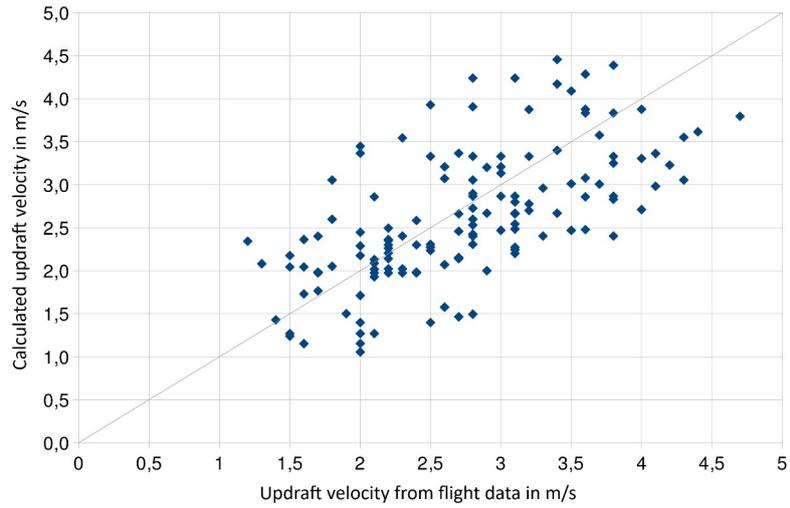

**Fig. 3:** Each point shows an evaluated updraft velocity from the flight data (X-axis) and contrasts it with the corresponding theoretical value from the thermal formula (Y-axis).

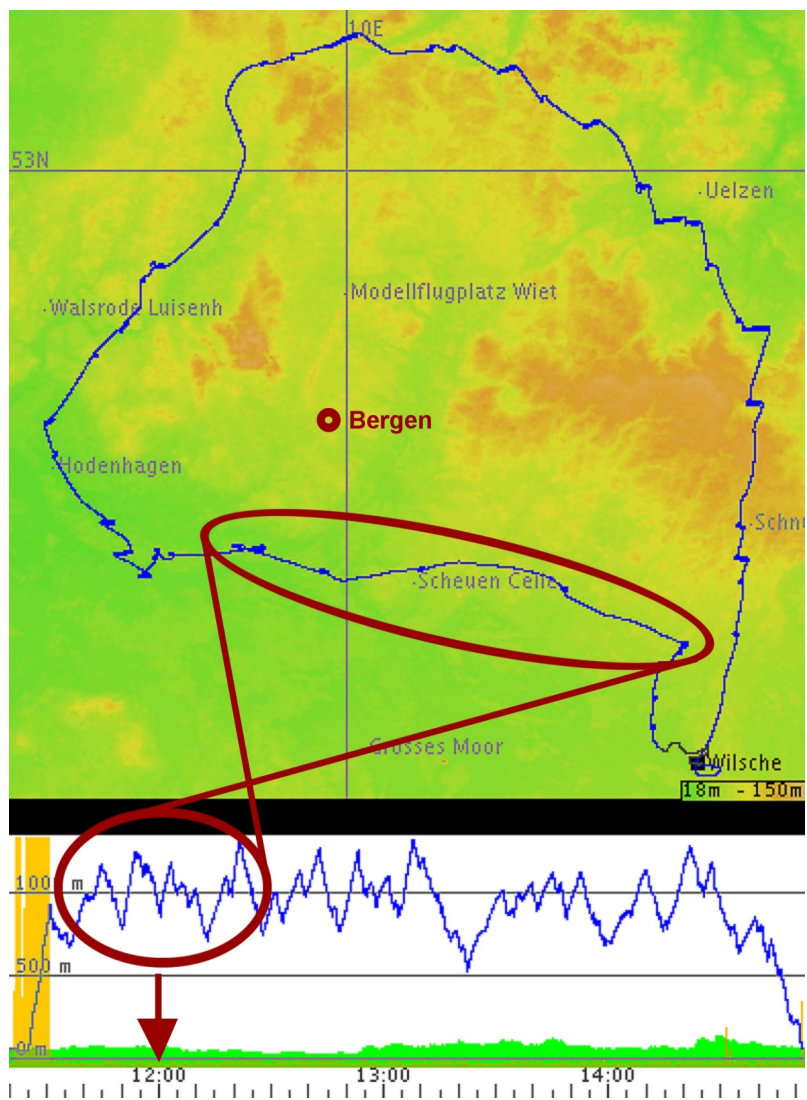

**Fig. 4**: Flight around Bergen (Germany) on July 7, 2016. Source: OLC.

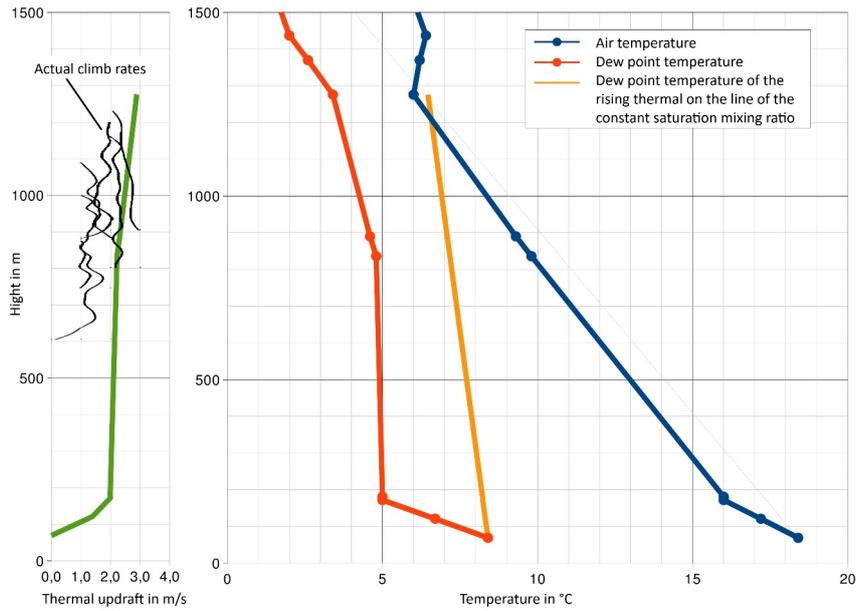

**Fig. 5**: Temperature readings from the Bergen (Germany) radiosonde on July 7, 2017, 12 UTC, with the calculated updraft velocity from them and the actual climb rates from Fig. 4.

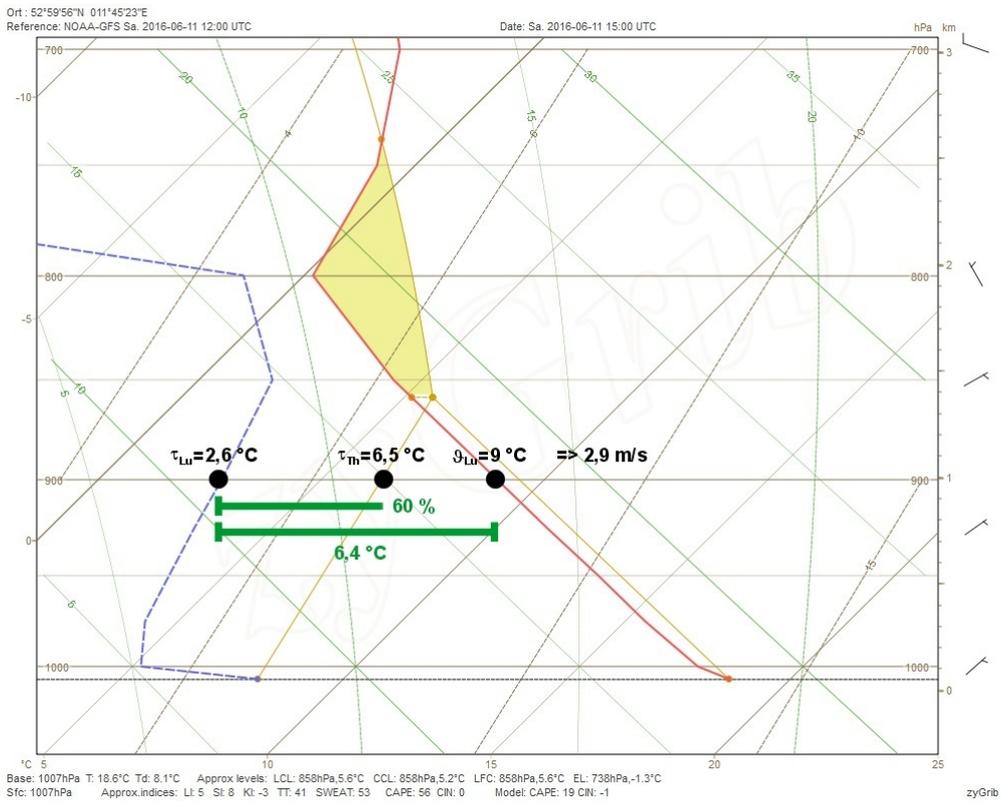

**Abb. 6**: Temp for Wittenberge (Germany) on 11 June 2016 and the thermal strength calculated from it. Source of the Temp: zyGrib.